*Probabilistic Spin Wave Computing with quasistatic magnetic inputs*

*Kirill A. Rivkin*

*RKMAG Corporation*

*254 Chapman Rd, Ste 208 #462, Newark, Delaware 19702*

*rivkin@rkmag.com*

*Spin wave computing device where an algorithm can be encoded by recording a corresponding magnetization pattern onto a hard magnetic material was previously proposed[1] and a particular implementation of a vector-matrix algorithm was demonstrated. In the present article we analyze the conditions allowing for implementation of complex algorithms which can combine multiple additive, multiplicative and conditional operators including logic expressions. Special attention is given to how the input data is provided. Rather than relying on a set of independent sources of the RF field as is common with the existing spin wave computing methods we demonstrate usability of more simple solutions using adjustable external quasistatic magnetic fields. We also show how for the given setup probabilistic switching of magnetic elements at elevated temperatures can be used to convert deterministic algorithms into a probabilistic form.*

**1. Introduction.**

In our previous work[1] we proposed a novel spin wave computing device which operates by creating a resonant mode with predetermined properties. It consists of two magnetic layers placed in close proximity to one another (Fig. 1): a "propagation layer", preferably made from a soft, low damping material (for example, permalloy), and a "bias layer", typically made from a hard magnetic material. Magnetization pattern recorded onto the bias layer subjects the propagation layer to a predetermined, non-uniform bias field, which in turn alters the scattering properties and therefore the spatial distribution of spin waves. Certain areas in the propagation layer identified as "inputs" and "outputs" respectively: to the former one applies external RF fields with amplitudes encoding the input values, from the latter one reads out the oscillation amplitudes as computation's results. The relationship between the two is defined by the resonant mode's spatial profile, which is controllable by the bias layer's magnetization pattern. It was demonstrated that as long as the linear mode of excitation is maintained[1] there is a number of algorithms which can be executed in such manner, including the vector-matrix multiplication. Using the perturbation theory we proposed a numerical method by which one can identify the magnetization pattern corresponding to a given set of matrix coefficients. Compared to more conventional approaches to spin wave computing utilizing domain walls, vortices or the physical shape of the propagation layer[2,3,4,5] there are distinctive advantages in terms of size (current technology allowing for the pattern modification on 10nm scale), cost and ability to rewrite the algorithm. Both in the present and previous manuscripts we consider the case where the bias layer is composed of nanodots (Fig. 1), but similar performance was observed with granular FeCo based films. It is even possible to make both the bias and the propagation layers from the same material (permalloy), relying on nanoparticles' shape anisotropy to maintain a stable magnetization configuration.

There are two key issues that need to be addressed for such setup to be practical. First has to do with how the input data is provided, which conventionally relies on a set of RF sources encoding the input values via different amplitudes. However, in reality adjusting the amplitude while maintaining the phase, i.e. synchronization between different sources, is difficult if not impossible. If there is a network of spin wave oscillators, it comes with an additional complication that changing the amplitude of an oscillator also impacts its operating frequency. There is also inherent device to device variation due to manufacturing precision. We demonstrate how such concerns can be addressed: both the inputs and the algorithm can be encoded as specific sections in the bias layer's magnetization pattern (Fig. 2), oralternatively (Fig. 5) the inputs can be provided as variations in the applied quasistatic (i.e. updated on a much slower scale compared to the mode's frequency) magnetic field. In both cases the RF field is supplied for the sole purpose of exciting the spin wave, carries no additional information and the requirements with respect to its amplitude and spatial distribution are greatly relaxed.

Second issue is that while it is pretty straightforward to adopt perturbation theory to describe magnetization patterns that can encode Vector-Matrix multiplication[1], it is of interest to address a more general question - which other operators can be encoded and how they can be combined together into algorithms. Relevant discussion is provided in the next Section of the manuscript.

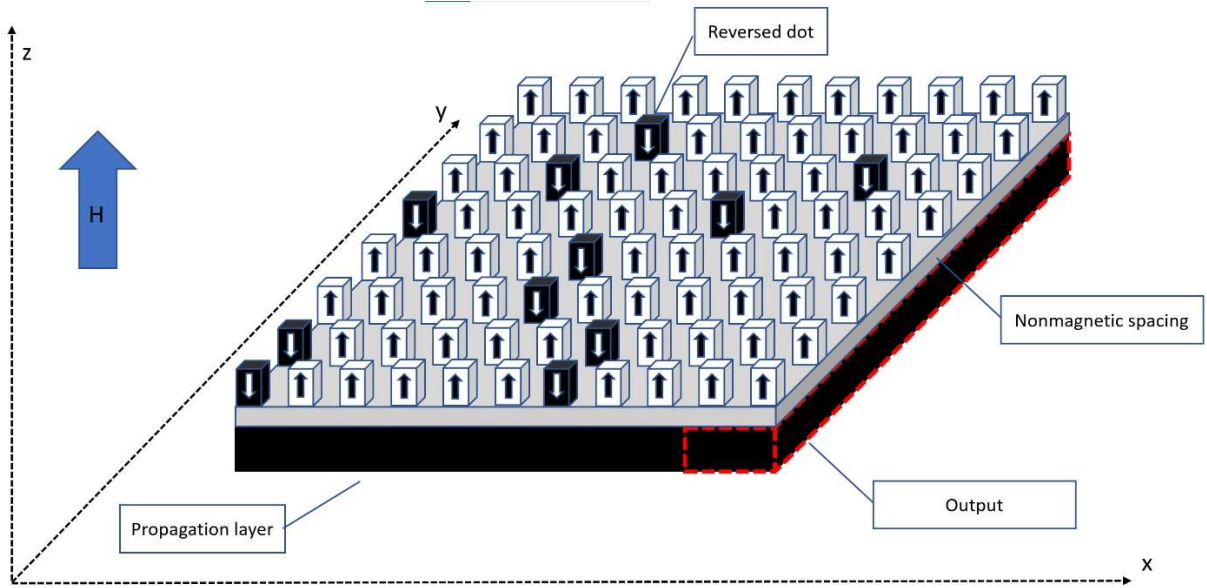

*Figure 1. Computing device considered in the present article. The propagation layer is a permalloy strip, 252x180x6nm in size, separated from the hard (bias) layer by 5nm non magnetic spacing. The hard layer is composed of 12x12x15nm FeCo dots, the interdot distance is 12nm. Most dots (depicted in white) have their magnetization aligned along the external magnetic field (applied along the z axis), but some are reversed (black dots) in order to encode the algorithm and possibly the input parameters. The output is read out as the amplitude of spin wave oscillations averaged over the rightmost 48x180nm section of the propagation layer.*

## 2. Additive and Multiplicative operators.

Particular implementation of the computational system being considered (Fig. 2) includes two magnetic layers. In the bias layer there are nanodots each of which can be reversed independently[6,7]. There is a uniform external field sufficient to predominantly saturate the propagation layer when the dots' magnetization is aligned along the external field, but not high enough to substantially affect the magnetization in the bias layer. In principle it can be applied in any direction, but in the present work we consider only the out of plane case (Fig. 1). As in the previous publication[1] we rely on linearized Landau-Lifshitz equation to produce a spectrum of resonant modes for the state with uniformly magnetized nanodots and then treat a magnetization reversal pattern as a perturbation. Using a bra-ket notation, $|V\rangle$ corresponds to the desired excitation mode of the perturbed case. An important assumption is that the output measurement can be approximated by a simple projection $\langle O|V\rangle = \int \boldsymbol{V}_m^{*(k)}(x,y,z) \cdot \hat{\boldsymbol{x}} H(x,y,z) d^3 x$ where $H$ is a nonzero constant only within the output area. In a general case there exists a correlation between the amplitude and the coordinate projection, but there is no one to one correspondence unless within the output area the magnetization is uniform and the modes are circular. Our justification is that the perturbation theory provides only the very basic estimation of the solution, for which purpose the assumption made appears to be practically sufficient.

Let us introduce two magnetization patterns in the bias layer which in the matrix form of the linearized Landau-Lifshitz equation[1] are represented by two perturbation matrices $A$ and $B$. Second order perturbation theory expresses result of the measurement $\langle O(A+B)|V\rangle$ as a perturbation of the $|V_0\rangle$ mode of the unperturbed system with a corresponding resonant frequency $\omega_0$:

$$\langle O|V\rangle = \langle O|V_0\rangle + \frac{\langle V_m|A|V_0\rangle}{\omega_0 - \omega_m}\langle O|V_m\rangle + \frac{\langle V_m|B|V_0\rangle}{\omega_0 - \omega_m}\langle O|V_m\rangle + \frac{\langle V_l|A+B|V_m\rangle\langle V_m|A+B|V_0\rangle}{(\omega_m - \omega_l)(\omega_0 - \omega_m)}\langle O|V_l\rangle$$

(Eq.1),

where summation is assumed over the indices *m, l*. Suppose we want the operators $A$ and $B$ to be additive, i.e. $\langle O|V(A+B)\rangle = \langle O|V(A)\rangle + \langle O|V(B)\rangle$. It is a straightforward point that such linearity requirement is fulfilled as long as both the quadratic term and the static bias $\langle O|V_0\rangle$ can be neglected, in which case (Eq.1) for arbitrary $A$ and $B$ becomes:

$$\langle O|V(A+B)\rangle = \frac{\langle V_m|A|V_0\rangle}{\omega_0 - \omega_m}\langle O|V_m\rangle + \frac{\langle V_m|B|V_0\rangle}{\omega_0 - \omega_m}\langle O|V_m\rangle = \langle O|V(A)\rangle + \langle O|V(B)\rangle$$

(Eq.2).

Assuming that the desired result $o = \langle O|V(A)\rangle$ of each operator is predetermined, the perturbation matrix must also satisfy[1] $A = \frac{o}{\langle O|V_m\rangle}(\omega_0 - \omega_m)|V_m\rangle\langle V_o|$. This expression is particularly useful when one needs to a rough guess in regards to the spatial distribution of $A$. Consider a more complex case: there is an existing set of additive operators $A$ satisfying (Eq.2) and we want to implement an additional operator $B$ such that:

$$\langle O|V(A+B)\rangle = b\langle O|V(A)\rangle \qquad \text{(Eq.3)},$$

where $b$ is a constant independent of the operator $A$. Numerically it is possible to solve the (Eq.1) directly[1], but to illustrate the physical meaning of such solution let us consider a simple case with only two modes having non-zero projections on the output area: $o_1 = \langle O|V_1\rangle$ and $o_2 = \langle O|V_2\rangle$ with corresponding frequencies $\omega_{1,2}$. For any operator $B$ satisfying the constraint:

$$\langle V_{1,2}|B|V_0\rangle = 0 \qquad (Eq.4)$$

the (Eq.1) becomes:

$$\langle O|V\rangle = \frac{\langle V_1|A|V_0\rangle}{\omega_0-\omega_1}o_1 + \frac{\langle V_2|B|V_1\rangle\langle V_1|A|V_0\rangle}{(\omega_1-\omega_2)(\omega_0-\omega_1)}o_2 = \langle O|V(A)\rangle\left(1 + \frac{\langle V_2|B|V_1\rangle o_2}{(\omega_1-\omega_2)o_1}\right) \qquad (Eq.5),$$

i.e. $b = 1 + \frac{\langle V_2|B|V_1\rangle o_2}{(\omega_1-\omega_2)o_1}$ for any additive operator $A$. The formula can also be inverted:

$$(\omega_1 - \omega_2)(b-1)\frac{o_1}{o_2}|V_2\rangle\langle V_1| = B \qquad (Eq.6),$$

providing the spatial dependence of $B$ as a function of the predetermined constant $b$. The purpose is to determine an approximate spatial distribution (dominated by the $|V_2\rangle\langle V_1|$ component[1]) and use it as an initial guess when solving numerically for the optimal magnetization pattern. It is straightforward to extend (Eqs.4-5) to the general case of multiple modes with non-zero projections $\langle O|V_m\rangle$.

Expressions (Eqs.1-6) are applicable to any linear excitation irrespective of its physical nature, i.e. optical, spin waves or other types of resonant modes. Main reason is why such general formulation is somewhat novel is that in most cases its of little use as it is near impossible to tailor the physical properties of the propagation media to satisfy the relevant constraints (for example, Eq.4). Magnetic systems however offer a number of enticing capabilities. There are stable sources of magnetic fields - magnetic media on which we have the ability to record near arbitrary shaped, stable magnetization patterns, producing magnetic fields with a spatial resolution on the order of 10-100nm. The application of the said fields alters the propagation properties and manipulates the perturbation matrix in (Eq.1). Spin wave modes of the unperturbed system in (Eq.1) can be obtained either numerically or analytically: the latter approach is useful when the propagation layer is large and uniformly saturated by the external field.

Joining together additive and multiplicative operators is straightforward since the relevant equations (Eqs.1-4) depend only on the spectrum of the resonant modes and thus for each operator one can devise an independent magnetization pattern which are then joined together in a modular fashion.

Unfortunately with algorithms which require interdependency of the operators (logic expressions included) simplified formulas like (Eq.5) are no longer sufficient and one has to solve the (Eq.1) or the full version of the linearized Landau-Lifshitz equation. Even for small samples (Fig. 2) estimating the system's response to each dot's reversal can be prohibitively expensive and it is important to devise a shortcut. For the AND operator a practically efficient path is to begin with a multiplication pattern which sums a set of even inputs with half of the inputs multiplied by "-1". It minimizes the projection $\langle O|V \rangle$ and suppresses the mode in the output area; using the corresponding pattern as a starting point we can then tweak it numerically as described below to produce a proper AND operator which returns low output amplitudes if the input values (operators) are strictly equal and large amplitude result otherwise (Table 1).

**3. Numerical modeling.**

Let us proceed with a numerical demonstration of a few simplistic algorithms. We assume the bias layer is made from a FeCo alloy with the saturation magnetization $\mu_0 M_s = 1.2$ T, magnetic anisotropy field 0.9T[6,7], damping coefficient $\beta = 0.06$, patterned as a periodic array of magnetic dots with a square cross-section (Fig. 1), 15nm thick, with center-to-center distance of 24nm and 12nm separation between the dots. The bias layer is separated from the propagation layer by 5nm non-magnetic spacing. The propagation layer is made from permalloy with the saturation magnetization $\mu_0 M_s = 1$T, damping coefficient $\beta = 0.01$, exchange stiffness 16 pJ/m. The overall system's size is 180x252nm and the corners of the propagation layer (Fig. 1) are aligned with the nanodots in the bias layer. The sample is very small compared to any practical system, but this allows for faster and more precise calculations, the purpose of the present article being a concept demonstration rather than analysis of a full size processor. The system is discretized into 3x3x3nm cubic cells and the dipole-dipole field is calculated by means of Fourier transform. There is a uniform external field along the z axis with the amplitude of 0.3T, noting that a larger sample would benefit from using a stronger field. Generally using field amplitudes comparable to or exceeding the propagation layer's demagnetization allows for better saturation of the propagation layer, improved tolerance to material imperfections and more reliable performance, while lowering the field so that the propagation layer is predominantly but not fully saturated increases the impact of each nanodot reversal and allows one to utilize more simple patterns. Free boundary conditions are employed. Solutions are obtained via the RKMAG package[8], capable of calculating magnetic equilibrium states, absorption spectrum and individual spin wave modes. For the output readout we average the amplitude in the rightmost 48nm section (Figs.1-2) of the propagation layer, the dots above this area remain magnetized parallel to the applied field. We reserve an area in the leftmost portion of the bias layer (Fig. 2) for the input data (patterns), further dividing it into two independent fields each responsible for encoding a single input: two fields each with 16 (four by four) nanodots, each encoding a single input value. The rest of the bias layer can encode some other operator $B$ (Fig. 2). The RF excitation can be applied uniformly to the entire sample, or be concentrated in any particular area; in the present case for the sake of convenience we use the RF field which is linearly polarized along the x axis and confined to the portion of the propagation layer directly underneath the bias layer's input segment.

For each computational operation it is important to define the error metric[1]. Given two input fields (Fig. 2) and N possible input values, for each pair of the input values $i,j$ let there is an expected output amplitude $C_{i,j}$. Since measuring absolute values of the amplitudes is difficult, we normalize the output amplitudes $C_{i,j}$ to the lowest possible value, which in the present case (Table 1) corresponds to both input fields carrying the input value identified as "1". Mean square root error metric $E$ denotes the difference between the observed amplitudes $\hat{C}_{i,j}$ and their intended values $C_{i,j}$:

$$E = \sqrt{\frac{\sum_{i=1}^{N}\sum_{j=1}^{N}\left(\frac{\hat{C}_{i,j}}{\hat{C}_{1,1}} - \frac{C_{i,j}}{C_{1,1}}\right)^2}{N^2}} \qquad (Eq.7)$$

The modeling procedure is as follows. First the equilibrium magnetization is calculated for the case of the nanodot array magnetized along the z axis and the linearized Landau-Lifshitz equation is used to obtain the entire spectrum of the resonant modes[8]. Next task is to produce the input operators (patterns): we demand that each corresponds to a specific input value and in the absence of other operators the output result is the sum of these values (Table 1). Using the (Eqs.1-2) we obtain a rough estimate of the corresponding magnetization patterns, and in a similar manner (Eqs.4-6) produce a pattern representing an additional multiplicative operator $B$. This by itself results in the values of the error metric $E$ as low as 0.1-0.2. Further improvement is obtained by modeling the changes resulting from the reversal of each of the nanodots individually, choosing the one demonstrating the largest gain and repeating the process, until after typically 3-4 iterations the error metric $E$ drops well below 0.01 for all cases considered in this manuscript. At this point precision of the micromagnetic modeling becomes a gating factor. Keeping the magnetization reversal pattern fixed and repeating the modeling of spin wave propagation (i.e. finding the equilibrium magnetic state including both soft and hard layers, applying the RF field, waiting for the steady state dynamic solution to form and then measuring the output amplitude) demonstrates that the repeatability of the error metric $E$ strongly depends on the chosen floating point precision - with double precision the values below 0.001 are not repeatable. This can be improved by expanding the output area which means averaging the output amplitude over greater number of discretized spins.

Another important consideration is how the applied RF field's frequency is selected. There are two considerations. First is computational precision: in order to minimize the error (Eq.7) and satisfy the constraints in (Eqs.2-5) there needs to be a considerable number of modes close to the operating frequency. Second consideration is making sure that the amplitude of the output signal is large enough to be detectable. In practical applications this is determined primarily by a detection mechanism chosen, including a post-processing channel and error correction mechanism. In a modeling exercise we can't account for such a complex system and have to opt instead for a shortcut. First we calculate the maximum oscillation amplitude at any frequency for the case of the uniformly magnetized bias layer. Since similar absorption peaks in nanoparticles are a common subject in published experimental studies we assume its amplitude to be sufficient for detection. Consequent magnetization reversals in the bias layer generate additional scattering points in the propagation layer and the maximum output amplitude decreases; we need to make sure it stays "comparable" to the peak amplitude of the uniformly magnetized case. Therefore, we demand that for the input operators by themselves the output amplitude is no less than 25% compared to the maximum absorption calculated for the case of the uniformly magnetized bias layer, and with an additional operator activated the amplitude should not drop below 15% of the said peak value. It can be noted that the extended resistance of similar systems to nonlinear behavior[1] should allow for relatively large RF fields (even as high as 100 Oe in case of a very small, permalloy based propagation layer) and in such case the output signal can be detected by almost any means.

| $I_1$ | $I_2$ | Additive | Modeled | AND | Multiplicative | Modeled | Using fields as inputs |
|---|---|---|---|---|---|---|---|
| 1 | 1 | 1 | 1 | 1 | 1 | 1 | 1 |
| 1 | 2 | 1.5 | 1.52 | 3.67 | 1.67 | 1.66 | 1.6 |
| 2 | 1 | 1.5 | 1.51 | 1.92 | 1.33 | 1.34 | 1.33 |
| 2 | 2 | 2 | 2 | 1.02 | 2 | 1.99 | 2 |

Table 1. Expected and modeled results using two input fields $I_1$ and $I_2$ and two possible input values (1 and 2), for the following three cases: input operators alone; input operators combined with a logical AND operator; input operators combined with a multiplicative operator (Eq.8); input values provided by quasistatic fields combined with a multiplicative operator encoded via the magnetization pattern (Fig. 5).

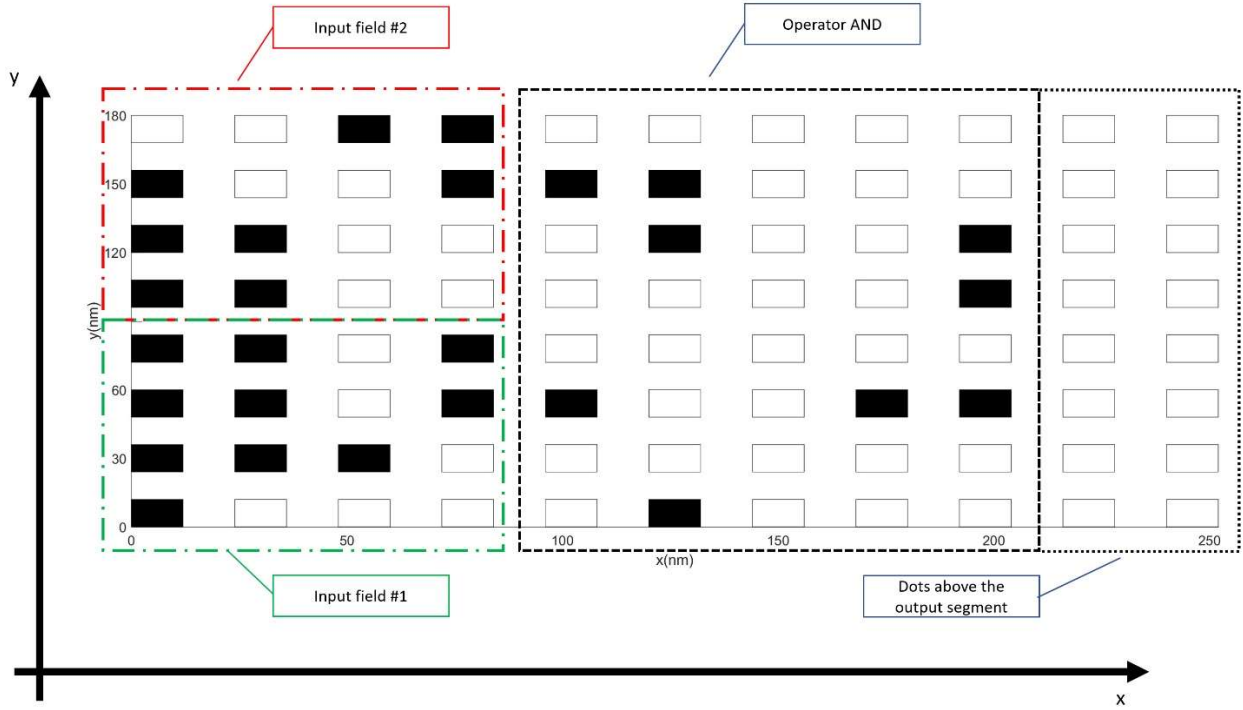

*Figure 2. Two input fields, field #1 containing the pattern corresponding to the input value 1, field #2 containing the pattern corresponding to the input value 2 ($I_1 = 1, I_2 = 2$), joined on the right by the pattern denoting logical AND operator. The dots with the magnetization aligned along the external field are white, with the reversed magnetization – black. Operating frequency f=2.8 GHz.*

Let us now demonstrate the behavior of a few specific algorithms. For the sake of simplicity we use only two additive input parameters: 1 and 2. If there are no additional operators the output should be the sum of the inputs, normalized to the lowest possible value ($I_1 = 1, I_2 = 1$, Table 1): when both inputs are 2 the output amplitude is supposed to be twice as large and a combination of 1 and 2 ($I_1 = 1, I_2 = 2$) yields 1.5 of the minimal output value. For the logical AND operator (Table 1) we require that when both input fields (Fig. 2) contain the same input patterns (values) the output amplitude is much lower compared to the case when the inputs are different. Finally for the multiplication operator we encode the following operation connecting the output amplitude $C$ with the input values in the first and second input fields, $I_1$ and $I_2$ respectively:

$$C = (b_1 \quad b_2)\begin{pmatrix}I_1\\I_2\end{pmatrix} = (1 \quad 2)\begin{pmatrix}I_1\\I_2\end{pmatrix} \qquad (Eq.8).$$

Normalizing to the smallest output (i.e. $I_{1,2} = 1$) yields the values in (Table 1). The input patterns remain the same in all three cases.

Magnetization pattern corresponding to logical AND operator as well as input values 1 (placed in the input field #1), and 2 (placed in the input field #1) is shown in (Fig.2); (Fig. 3) demonstrates various combinations of the input values and the resulting amplitudes of magnetic oscillations in the propagation layer.

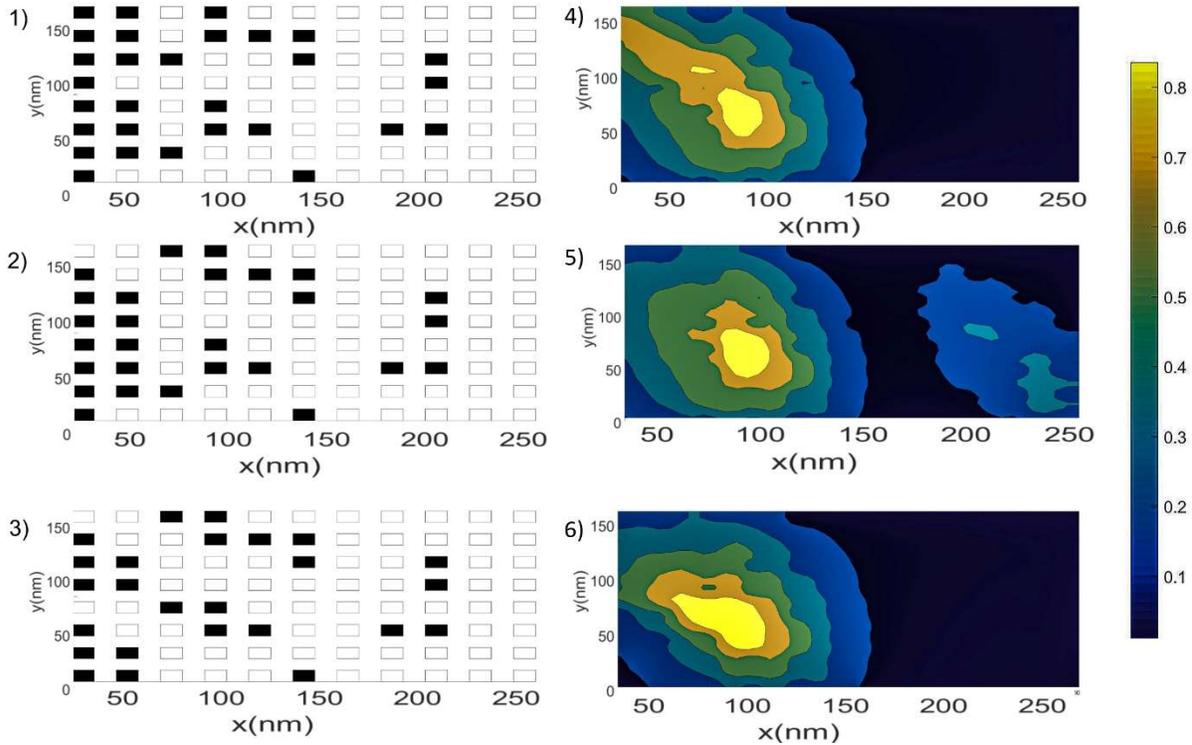

*Figure 3. Three cases illustrating the application of logical operator AND (Fig.2). (1-3) represent the encoded patterns corresponding to different inputs ($I_1 = 1, I_2 = 1$ for 1; $I_1 = 1, I_2 = 2$ for 2; $I_1 = 2, I_2 = 2$ for 3), (4-6) the resulting magnetic oscillation amplitudes in the propagation layer, normalized to the maximum observed for all cases. Operating frequency f=2.8 GHz.*

The amplitude in the output segment of the propagation layer (Fig. 3.4-6) behaves as intended: if the patterns in the input fields are mismatched (Fig.3.2) the output amplitude visibly increases (Fig.3.5).

The operator AND can be substituted by a multiplication operator (Eq.8) by rewriting the corresponding section in the magnetization pattern (compare Fig. 2 and Fig. 4) while keeping the input operators representation the same (Fig. 4). The output amplitude is again normalized to the case $I_1 = 1, I_2 = 1$ and a good match is observed between the modeled and predetermined values (Table 1).

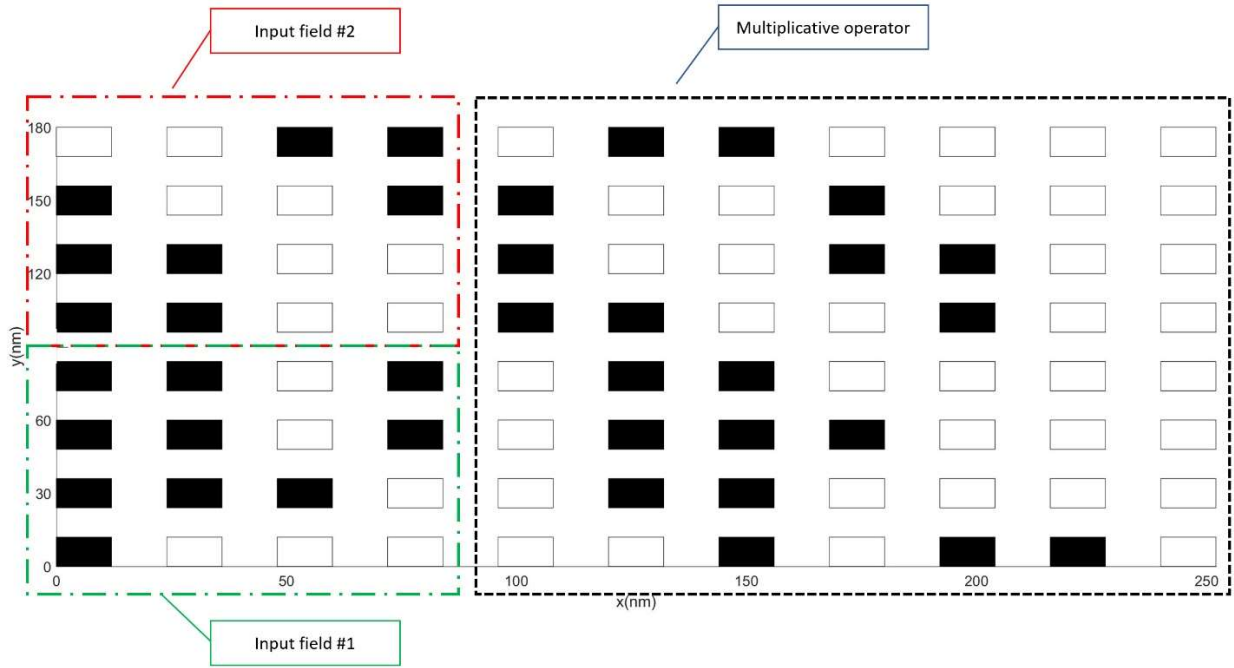

Figure 4. The input values $I_1 = 1, I_2 = 2$ are joined on the right by the pattern encoding the multiplicative operator (Eq.8). Operating frequency f=2.8 GHz.

**4. Quasistatic field as inputs and probabilistic algorithms.**

Reversing magnetization patterns in nanodots or granular films can be a more practical approach when compared to using independent synchronized sources of the RF field. However, magnetization reversal also involves a number of operational constraints. Using a magnetic memory inspired approach (for example, by placing spin torque based memory cells in the input fields of the bias layer) enables a fast random access, but requires expensive wafer based manufacturing and limits the minimal size of individual switchable elements. A rewritable media like the one used in hard drives can be ideal for a long term, seldom updated algorithms but can be inefficient if the input data needs to be changed on a regular basis.

To address this challenge we can point out the fact that (Eqs.1-6) are applicable to any perturbation that impacts the magnetization dynamics. One of the simplest to implement possibilities is to add a set of external quasistatic magnetic fields whose amplitudes and or spatial distributions denote the input values. A computational operator is still encoded via the magnetization pattern, while external magnetic fields can be produced by a simple source such as a current carrying wire placed in the propagation layer's vicinity.

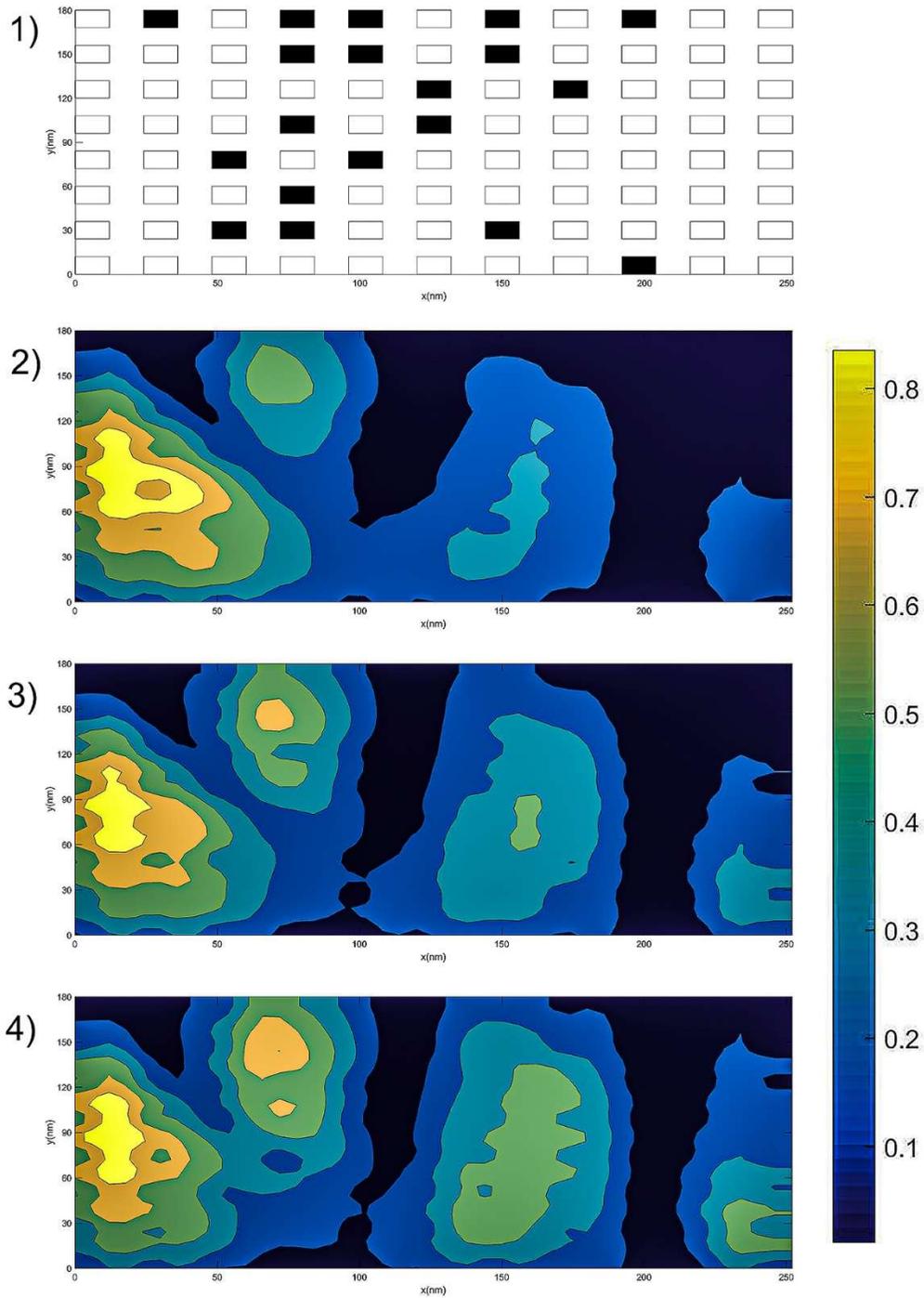

*Figure 5. Multiplicative operator (Eq.8) encoded via the magnetization pattern (1) with inputs provided by static uniform magnetic fields aligned with z axis and confined to the input fields #1 and #2 identified previously in (Fig. 2). The amplitude 300 Oe denotes value 1, 900 Oe – value 2. (2-4) present the amplitudes*

*of the excited mode for the case of inputs $I_1 = 1, I_2 = 1$ (2), $I_1 = 1, I_2 = 2$ (3), $I_1 = 2, I_2 = 2$ (4). Operating frequency f=2.8 GHz. Operating frequency f=5.88 GHz.*

In the most simple case (Fig. 5) different values of the inputs can be presented by different amplitudes of the applied fields. These fields are confined to the areas in the propagation layer directly underneath the segments in the bias layer (Input fields #1 and #2) previously used to provide the inputs (Fig. 2), aligned along z axis, the amplitude 300 Oe denotes the value 1, 900 Oe – the value 2. The multiplicative operator (Eq.8) can then be encoded via the magnetization pattern shown in (Fig. 5.1), and the results are a good match for the intended values (Table 1, Fig.5.2-4): changes in the amplitude of magnetic fields applied to the "input areas" successfully modifies the mode's behavior in the output segment. It can be shown that for more realistic scenarios involving a significant number of both input fields and input values it is imperative to encode each input through a combination of at least two spatially non-uniform field sources so that the spatial distributions in (Eqs.1-6) can be more accurately satisfied. Even then the input interface based on quasistatic magnetic fields remains simpler, cheaper and often faster compared to the RF based alternatives.

Temperature is another environmental factor which affects the magnetization dynamics and thus can be used to alter the behavior of a computational system. Background temperature was previously shown[1] to be a precision limiting factor whose impact can be mitigated to an extent by an appropriate adjustment of the magnetization pattern, in the same manner as it is possible to account for imperfections in magnetic material and geometry. A more interesting case is when a temperature profile is applied to a specific portion of the bias layer for a duration sufficient so that the affected nanodots can reverse their magnetization with a given probability. (Eqs.1-6) demonstrate that the scale of alterations affecting the algorithm depends on the wavelength being excited, which in turn depends on the system's size and other factors. For the present case the reversal of even a few neighboring dots is sufficient. The location, peak temperature and duration of application can be selected in such a way so that the algorithm parameters are affected rather than the nature of the algorithm itself - for example, the multiplication in (Eq.8) attains probabilistic multiplication coefficients but remains a multiplication, and also the probabilities corresponding to different parameters are comparable to one another.

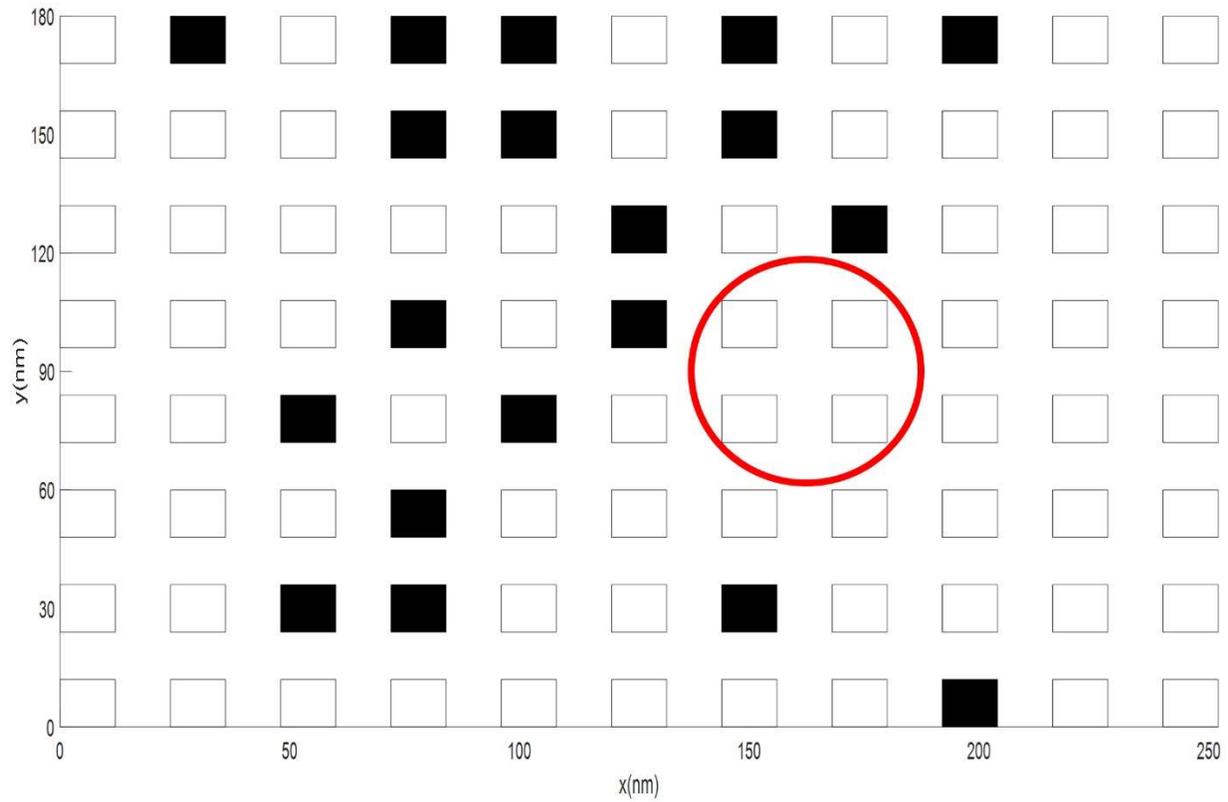

*Figure 6. Magnetization pattern for the multiplication operator with magnetic field based inputs (same as in Fig. 5) with a Gaussian thermal profile (1150K peak temperature, 20nm half width, 15ns duration) applied at the center of the red circle.*

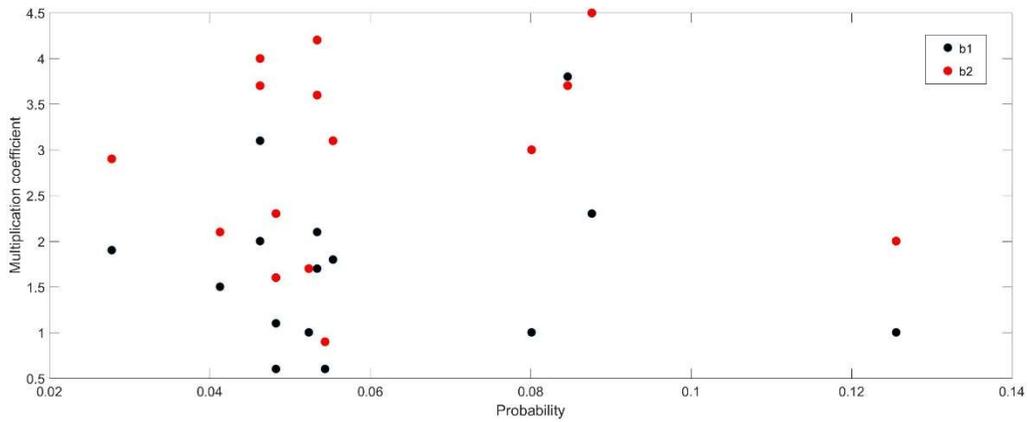

*Figure 7. Multiplication coefficients $b_{1,2}$ in (Eq.8) altered by the application of the thermal profile in (Fig. 6).*

As before, we consider the case when the computational system operates at zero temperature (condition chosen purely for the modeling convenience), except for a pulse with a peak temperature of 1150K, Gaussian spatial distribution with 20nm half width and duration 15ns applied to a specific portion of the bias layer (Fig. 6). The switching probability is calculated given the following assumptions: only four dots within the Gaussian profile are affected, each of the dots is subject to a combination of constant mean field produced by the unaffected dots and a dynamic mean field due to the magnetization averaged for each of the other three dots impacted by the temperature profile. This allows for each of the four dots to be modeled independently from the rest of the system: using micromagnetics to obtain correct estimation of the magnetization dynamics at such high temperature requires substantially smaller (1x1x1nm) discretization and a shorter integration timestep. After the thermal pulse is instantaneously activated and then switched off (using an exponential decay and risetime would be more physical but would do little to alter the conclusion) the magnetization is allowed to relax for 10ns. By repeating the modeling a probability value is assigned to each of the possible combinations, i.e. between zero and all four magnetic dots being reversed. Then for each of these magnetization patterns the output amplitude is computed for all four combinations of the input values (two input fields with two input parameters each); the result is fitted to (Eq.8) to make sure that the error metric remains well below 0.1 value, i.e. the algorithm remains a multiplication with the new values of the coefficients $b_{1,2}$ which are fitted using the modeling results with a step size of 0.1. The step size is large enough to group together physically similar results, i.e. those with the same reversed dots but slightly different magnetizations obtained at the end of 10ns relaxation period. Fitted multiplier coefficients with their respective probabilities are shown in (Fig. 7).

High temperature results (Figs. 6-7) are meant as an illustration of the system's capability for probabilistic computing. Admittedly the modeling procedure chosen is simplistic. On the one hand the dots are small enough so that the brute force micromagnetics can be both accurate and efficient, while the distances between the dots limit the importance of interdot interactions allowing for a mean field approximation. However a number of assumptions made are artificial though arguably not unphysical. Most important ones are that there is no thermal profile in the propagation layer and thus the collective switching phenomena considered at length by the spin ice community[9] can be neglected.

**5. Conclusions.**

Engineering spin wave properties by means of a complex bias field produced by a magnetization pattern recorded onto a hard media is a powerful and versatile tool capable of implementing a diverse range of computational algorithms. In the previous publication[1] we demonstrated performance exceeding that of conventional electronics in terms of speed, size and power consumed. This is not unexpected when algorithms are implemented directly through physical phenomenon (i.e. modern form of analogue computing), and while propagation speed of spin waves is below that of optical waves, typical operating frequencies are still in 1-50 GHz range.

First contribution of the present work was to extend the perturbation theory approach[1] to establish three different types of operators (additive, multiplicative and conditional) and demonstrate how each type can be implemented by means of magnetization patterns which can be then joined together into a single computing program. The inputs can be treated as operators and also implemented via corresponding magnetization patterns: unlike typical von Neumann architecture the proposed system does not have to differentiate between algorithm and data storage. When "on the fly" modification of either the inputs or the algorithm is needed, it is possible to modify the computation by changing the amplitude or spatial distribution of external magnetic fields. This greatly reduces the system's cost and complexity and can be a versatile solution when one needs to enable a communication between a magnetic processor and a device which employs electrical signals, i.e. currently existing electronic components.

It is also possible to convert a deterministic algorithm into a probabilistic one by utilizing an intrinsically probabilistic nature of magnetization reversal. Applying a thermal profile to a specific portion of the bias layer allows one to selectively alter the probability distribution of specific computational parameters.

**6. Data availability.**

The data that support the findings of this study are available from the corresponding author upon reasonable request.